\documentclass[preprint,12pt,authoryear]{elsarticle}

\usepackage{amssymb}
\usepackage{amsmath}

\usepackage{url}

\journal{Journal of Informetrics}

\begin{document}

\begin{frontmatter}

\title{Regional profile of questionable publishing}

\author[1,2]{Taekho You} 

\affiliation[1]{organization={Institute for Social Data Science, Pohang University of Science and Technology},
            addressline={77 Cheongam-ro}, 
            city={Pohang},
            postcode={37673}, 
            state={Gyeongsangbukdo},
            country={Republic of Korea}}

\affiliation[2]{organization={Center for Digital Humanities \& Computational Social Sciences, Korea Advanced Institute of Science and Technology},
            addressline={192 Daehak-ro}, 
            city={Yuseong-gu},
            postcode={34141}, 
            state={Daejeon},
            country={Republic of Korea}}

\author[3]{Jinseo Park}

\author[3]{June Young Lee}

\affiliation[3]{organization={Center for Global R\&D Data Analysis, Korea Institute of Science and Technology Information},
            addressline={66 Heogi-ro},
            city={Dongdaemun-gu},
            postcode={02456},
            state={Seoul},
            country={Republic of Korea}}

\author[4]{Jinhyuk Yun\corref{cor1}}
\ead{jinhyuk.yun@ssu.ac.kr}

\affiliation[4]{
            organization={School of AI Convergence, Soongsil University},
            addressline={369 Sangdo-ro}, 
            city={Dongjak-gu},
            postcode={06978}, 
            state={Seoul},
            country={Korea}}
            
\cortext[cor1]{Corresponding author}

\begin{abstract}
Countries and authors in the academic periphery occasionally have been criticized for contributing to the expansion of questionable publishing because they share a major fraction of papers in questionable journals. On the other side, topics preferred by mainstream journals sometimes necessitate large-scale investigation, which is impossible for developing countries. Thus, local journals, commonly low-impacted, are essential to sustain the regional academia for such countries. In this study, we perform an in-depth analysis of the distribution of questionable publications and journals with their interplay with countries quantifying the influence of questionable publications regarding academia's inequality. We find that low-impact journals play a vital role in the regional academic environment, whereas questionable journals with equivalent impact publish papers from all over the world, both geographically and academically. The business model of questionable journals differs from that of regional journals, and may thus be detrimental to the broader academic community.
\end{abstract}




\end{frontmatter}



\section{Introduction}

Questionable journals and publishers gradually expand their academic market share and influence. They employ profit-driven models and attract researchers who require imminent outcomes. For instance, they exploit insufficient peer review process~\citep{bohannon2013s} and it introduces problems in academia, such as plagiarism~\citep{Owens2019JNS} and misinformation~\citep{beall2016medical}. These problems will decrease academic credibility~\citep{west2021misinformation}. To protect academia, academic stakeholders, including governments, indexing services, and authors, have issued forewarnings against their misconduct and provided warning lists ~\citep{beall2016essential,anderson2017cabell,kakamad2019kscien}. Despite these efforts, an increasing number of authors are involved in questionable publication.

It is necessary to understand whether the publication trends in questionable journals differ in terms of authors. Questionable publishing tends to be recognized as a problem for less developed countries because these countries have accounted for the majority of publications in questionable journals~\cite{shen2015predatory,sterligov2016riding,machavcek2017predatory}. Indeed, studies found that authors in these countries have occupied a major fraction of publications in questionable journals~\citep{xia2015publishes,nwagwu2015penetration}, even though these questionable journals advertise themselves as international journals; their potential strategy to attract authors due to a requirement for international publication~\citep{shen2015predatory}. On the other side, the governmental evaluation system that relies on the quantitative evaluation, \textit{e.g.}, simple counting the number of papers, is sometimes claimed as the origin of such phenomena, which leads authors to be easily published in questionable journals~\citep{raghavan2014predatory,omobowale2014peripheral,quan2017publish,demir2018predatory,kurt2018authors}.

The social stratification in academia can be another reason for publishing in questionable journals. Authors have published in questionable journals after experiencing failures to publish in reputed journals~\citep{demir2018predatory}. Studies have found a concentration of citations and publications among majority authors due to bias against minorities in terms of gender, race, ethnicity, and country~\citep{huang2020historical,nielsen2021global,sekara2018chaperone,kozlowski2022intersectional}, suggesting that academia is becoming more unequal. In addition, the spatial localization of knowledge and technology influences scientific and technological innovation~\citep{FELDMAN2010381}. Academic collaboration has emerged more frequently in top-tier universities~\citep{jones2008multi}. The economic scale of countries is another factor, as modern scientific practice requires substantial investment, making it more challenging for developing countries~\citep{miao2022latent}. Consequently, researchers in less developed regions may be compelled to choose questionable journals, despite their low academic impact. These publication trends suggest that questionable publishing meets a certain demand from less developed countries; therefore, criticisms of their contributions to questionable publications may seem unfair from their perspective. However, further in-depth analysis of publication trends in terms of authorship remains necessary.

Neither approach may have a complete understanding of the role that questionable journals play in academia. There are criticisms of questionable journals, but this is because it is difficult to reject without solid evidence that they are a necessary evil or that they supplement the limits of the current journal system to some degree~\citep{machavcek2022predatory}. To comprehend the role of questionable journals in these two contradictory perspectives, especially focusing on the local academia, we believe that it is required to examine the relationship between the publication country and the authors' affiliated country, as opposed to the analysis of publication frequency or publication rate by the country that has been conducted thus far. If questionable journals serve the local academia, contributions from neighboring nations should be abundant. In this study, we examine a paired profile of questionable and unquestionable journals with comparable impacts. We considered journals named in Beall's list~\citep{BeallsWeb} as questionable and unquestioned journals as a comparison set, which shows similar chracteristics with each corresponding questionable journal, yet not accused by Bealls's list~\citep{you2022disturbance} (see Section~\ref{Methods} for details). We quantitatively determined the regionality of journals dominated by a single country and compared the regionality of questioned and unquestioned journals. Based on the findings, we observe that questionable journals contributed from more various and non-adjacent countries than unquestioned journals; therefore, questionable journal plays a very limited role in sustaining local academia and expansion of such journals will not help the healthy academia in developing countries.

\section{Methods}\label{Methods}

\subsection{Data}

In this study, we compare the affiliation information between authors and journals. We use the country information from the affiliations with each country's gross domestic product (GDP) by analyzing the scholarly publication data from SCOPUS. We first extracted 20,864,247 papers, 6,971,029 authors, and 27,938 journals from the January 2019 dump of Scopus CUSTOM XML DATA. We then restricted data dating from 2010 to 2018 and having affiliation information, which the author's data contains 906,610 affiliations with 177 countries. To reduce the potential statistical error, \textit{i.e.}, fluctuation from the low number of publications, we excluded journals that published less than 30 publications during the timeframe.

\subsection{Selection of questionable and unquestioned journals}

To identify the questionable and unquestioned journals, we employ the pre-complied list from our own previous study based on Beall's list~\citep{you2022disturbance}. We cross-checked their ISSN and name between the SCOPUS dataset and the list, and excluded journals that publish less than 30 articles per year. In total, we found 766 questionable journals along with their 1,293 unquestioned counterparts. One questionable journal can be matched with multiple unquestioned journals because we assigned an unquestioned journal for each ASJC category of a target questionable journal.

We defined comparative unquestioned journals, matched with questionable journals, using three criteria from our previous study~\citep{you2022disturbance}: (1) similar journal impact, (2) same subject area classification (ASJC) in Scopus, and (3) a similar amount of annual publications. First, we limited the candidate journals to those belonging to the same subject area (ASJC). Second, we divided journals into three groups according to their annual publication volume and limited the candidate journals to those within the same group. Finally, we selected unquestioned journals with a similar journal impact. Note that we computed the journal impact using the Journal Citation Reports' Impact Factor methodology, but calculated it with the SCOPUS dataset. As a result, we yield 1,293 unquestioned journals since one questionable journal belongs to multiple subject areas. We should note that the purpose of the comparison focuses on figuring out the overall patterns of questionable journals and publishers rather than revealing differences between two individual journals. To compare the tendencies of general journals without considering journal impact, we used another set called other journals, which consisted of all 25,879 Scopus journals not included in the two above sets.

\subsection{Allocating countries to journals}

When deriving country profiles for both journals and papers, it is essential to assign the correct country to each. We assume that the first author plays the most significant role in the publication; therefore, we consider the paper to belong to the first author's country. Since a single author may have multiple affiliations, we set the sum of the author's countries as given equal weight as one: as an illustrative example, $2/3$ of country A and $1/3$ of country B are assigned to the paper if the first author has two affiliations with country A and one affiliation with country B. Some papers have equally contributing first authors, yet we consider a single author who appears first in the author list to be the first author regardless of their contribution markups. Additionally, the corresponding author may play a central role in the publication process. We found that $91.3\%$ of papers list the same country for both the first and corresponding authors; thus, considering the corresponding author as the first author should not significantly change the results. In the meantime, we employ the country of the journals belonging to as listed in the Scimago Journal \& Country Ranking~\citep{sjr}; we hereafter call it as ``publishing country''.

To define neighboring countries for a given country, we construct two types of networks: i) by the geographical distance and ii) by academic distance as the inverse number of co-authored papers between countries. The geographical distance between two countries is measured by the haversine distance between their respective centroids~\citep{centroid}. We also count the number of co-authored papers as the number of papers whose affiliations include both countries. Here, we consider papers with two or more authors solely to exclude the case that a single author has multiple affiliations with different countries. We then select five topologically, geographically, and academically nearest countries as the neighboring countries regarding a given publishing country.

\section{Results}

\subsection{Overestimation of questionable contributions from less developed countries}

\begin{figure}
    \centering
    \includegraphics[width=\textwidth]{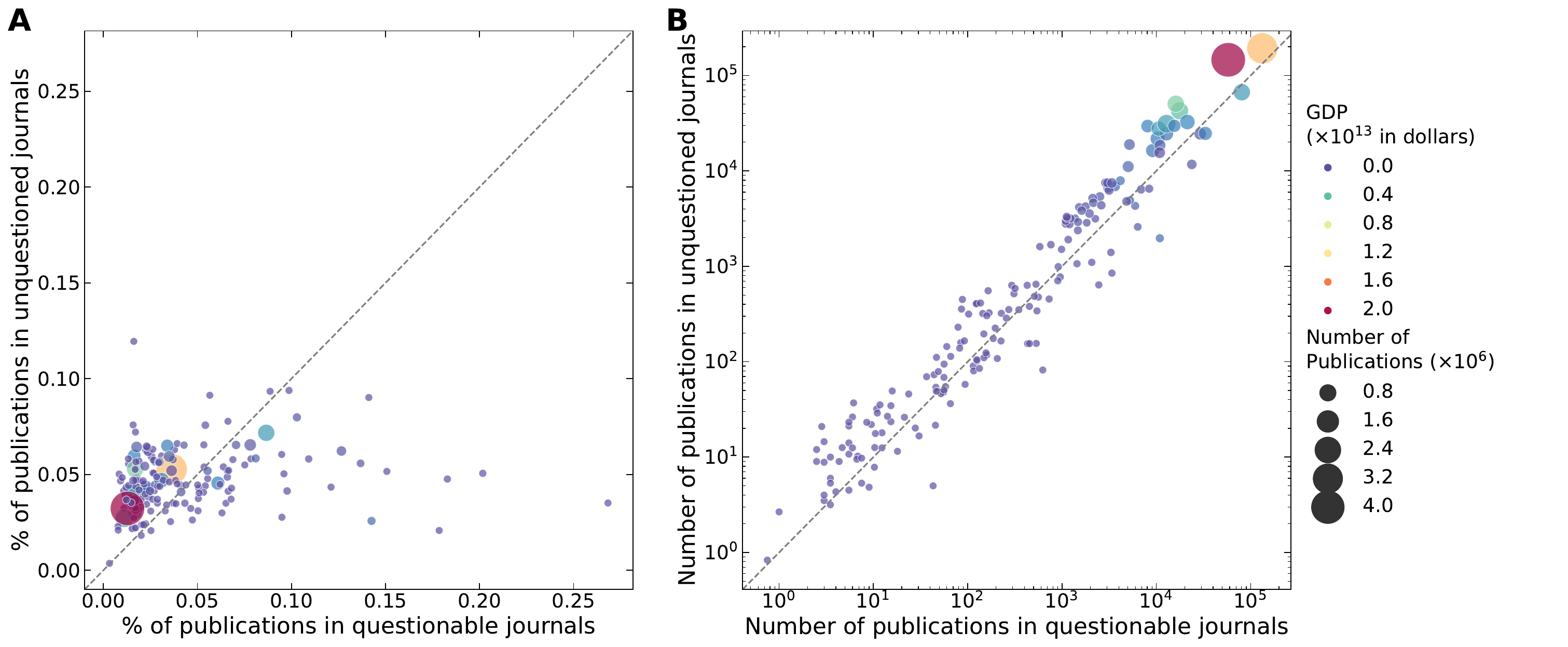}
    \caption{\textbf{Country-specific publication statistics for questionable and unquestionable journals.} The diameter of the circle represents the country's total number of publications, whereas its hue represents the GDP in 2018. Countries that published fewer than 100 papers are excluded from the visualization.
    \textbf{(A)} Countries' proportion of questionable and unquestionable publications among their entire publications. Countries with a high GDP publish a lower proportion of questionable journals, whereas countries with a low GDP publish a greater proportion of questionable journals. 
    \textbf{(B)} The number of questionable and unquestioned publications by country. Countries with a high GDP publish more in both questionable and unquestionable journals than those with a low GDP. The majority of countries are arranged in a diagonal line, indicating that they generate a comparable number of publications in both questionable and unquestionable journals.}
    \label{fig:fig1}
\end{figure}

According to former studies~\citep{machavcek2017predatory}, the incidence of questionable publications is higher in less developed countries than in developed countries (Figure~\ref{fig:fig1}A). We discovered that 122 countries out of 177 publish less than 5\% of their publications in questionable journals. Meanwhile, low-GDP countries publish more than 10\% of their publication in questionable journals. These countries are mainly located in Southeast and Middle Asia, and their GDPs are less than 5\% of that of the country with the highest GDP. The result is consistent with previous studies indicating that less developed countries are producing an increasing number of questionable publications.

We discovered, however, that the contribution of less developed nations to questionable publishing seems to be overestimated. The Pearson correlation between the number of publications in questionable journals and the GDP is $0.76$ (Figure~\ref{fig:fig1}B), while the correlation between the rate of questionable publications and the GDP is $-0.07$ (Figure~\ref{fig:fig1}A). Although the proportion is low in comparison to the total number of published papers, the high-GDP countries produce a larger number of questionable publications because they produce more publications in total (Figure~\ref{fig:sm_publication_size}). Countries publishing more than 10\% of their publications in questionable journals occupy 9\% of entire questionable publications but only 2\% of all academic publications during the same period. In contrast, the United States, the highest-GDP country, is responsible for 20\% of all scholarly publications and 9\% of all questionable publications. This finding suggests that low-GDP countries have a limited impact on the growth of questionable publications. Meanwhile, despite a small publication rate, countries with a high GDP produce more questionable publications. According to previous studies~\citep{beall2012predatory,memon2019revisiting}, most questionable journals employ the open-access model, which indicates that high-GDP countries contribute more to growing questionable journals in terms of both publication volume and profit. In conclusion, the disparity in the denominator, \textit{i.e.}, the total number of publications in the country, considerably contributes to the overestimation of questionable publishing in countries with a low GDP.

Note that the majority of countries are distributed along the diagonal line $y\sim x$ (Figure~\ref{fig:fig1}B). Even though we do not consider the country in the filtering process for unquestioned journals, we discovered that the country produces a similar number of questionable and unquestioned publications. The result merely indicates that there is no disparity in the selection of journals at the country level. Therefore, although authors from less developed countries have a greater likelihood of publishing in questionable journals, their contribution to such publications may not be significant, regarding total impact.

\subsection{Regional distribution of questionable journals associated with neighboring countries.}

\begin{figure}
    \centering
    \includegraphics[width=\textwidth]{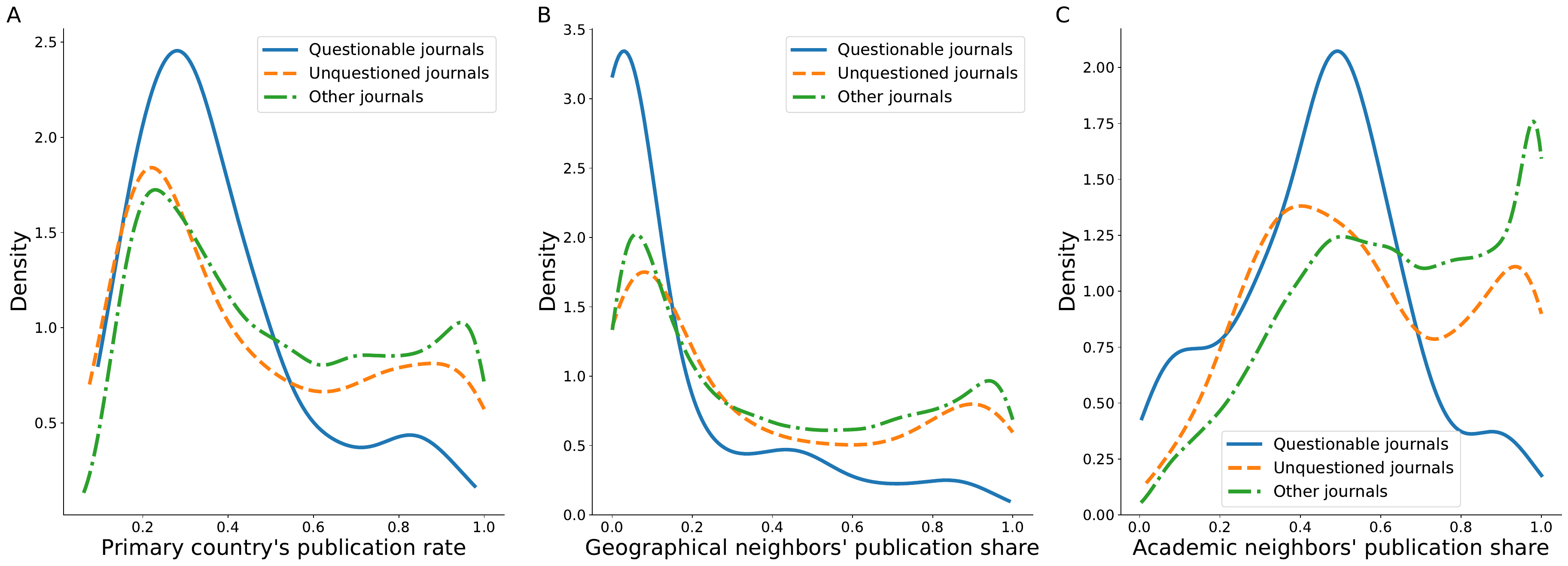}
    \caption{\textbf{Distribution of journals based on the proportion of publications from three different definitions of neighboring countries.} 
    \textbf{(A)} The publication rate of the primary country in the journal. 
    Here, the primary country is the first author's country that has the largest share of a given journal.
    \textbf{(B)} Publication proportion for geographical neighbors.
    \textbf{(C)} Publication proportion for academic neighbors.
    To minimize noise, each figure is rendered using the KDE plot.}
    \label{fig:fig2}
\end{figure}

According to the aforementioned observations, the impact of countries with a high GDP is frequently overlooked in questionable publications regarding their significant contributions. Nonetheless, another noteworthy aspect is the location where journals are published (Figure~\ref{fig:sm_journal_size}). There is a strong correlation between non-questionable journals and the GDP of the country of publishing ($0.72$ for unquestioned journals and $0.75$ for other Scopus journals). Thus, the greater a country's economic strength, the more scientific journals it publishes. However, the questionable journals have a weaker Pearson correlation with the country's GDP of only $0.30$.

Authors select journals based on their empirical and local information, including academic reputation, quantitative metrics, editor, and personal preference~\citep{frank1994authors}. Authors tend to perceive that international readership has been influenced by the journal's title~\citep{jamali2024country}. Quantitative indicators are likely to be low for these regional journals because their readership is less extensive than that of international journals, which have potential readers from all over the globe~\citep{abramo2016effect}. Nonetheless, because these regional journals mainly tackle issues of regional significance, they play a crucial role in the academic ecosystem of the region, which is not reflected in a quantitative indicator.


From this perspective, the concept of regionality may provide additional insight into the current trend of questionable publishing. If questionable journals address significant issues in a certain region, these journals will play an important role as new alternatives within the regional academic community. This possibility can be inferred from the fact that questionable journals have a larger market share in developing nations. To test this hypothesis, we first calculate the proportion of the most-published countries in a given journal. The simple calculation demonstrates that the majority of questionable journals are not occupied by a single country (Figure~\ref{fig:fig2}A). Here, we define the primary country as the author's country that has the largest share of a given journal. Questionable journals have a lower share of the primary country ($0.38$) than unquestionable ($0.47$) journals and other Scopus journals ($0.51$) on average. To say the dominant country is the country that publishes more than $60\%$ of papers in a single journal, \textit{i.e.}, more than a half of publications are occupied by a single country, only $15\%$ of questionable journals have dominant country while $34\%$ of other types of journals have one. This behavior can be understood by another regionality reflected in the written language of papers from the primary countries. We discovered that 148 unquestionable journals accept articles in non-English, whereas only seven questionable journals do so. In other words, while a significant proportion of unquestioned journals accept non-English articles for publication in their respective local academia, questionable journals with comparable quantitative metrics did not.

The difference between the primary and publishing countries is a possible explanation for why questionable journals are less dominated by the primary country. We analyze the primary country's publication rate with its relationship to the publishing country of the journal (Figure~\ref{fig:sm_dominant_country}). 53\% of all journals display the same country of publication and primary country, whereas 47\% do not. For journals with identical primary and publishing countries, the publication share of primary countries is distributed nearly uniformly, with a peak near 1. There are 30.7\% of these journals where the primary country generates more than 80\% of all publications. In contrast, if the two countries are dissimilar, the publication proportion of primary countries in a journal is reduced. We discovered that only 22.3\% of questionable journals are published in their home country, suggesting that the primary country has a lower publication rate in the remaining 77.7\% of questionable journals. In short, if the primary country is identical to the publishing country, the primary country may dominate more. Since the primary country is different from the publishing country of many questionable journals, the primary country's publication rate is low.

However, the relationship between the primary and publishing country cannot completely explain regionality. For instance, a substantial number of journals have high publication rates in primary countries that are not the same as the publishing country. The rapid expansion of mega-publishers, to which an increasing number of journals now belong, is one possible explanation for this circumstance. Citation indexes commonly attribute the location of publishers as the country of publication, although they are representative of a particular regional academic community. As an illustration, the ``Korean Journal of Chemical Engineering'' is the official publication of the ``Korean Institute of Chemical Engineers''; however, SJR lists its publishing country as the United States because it is published by \textit{Springer New York}. In addition, the local academic community may encompass neighboring nations that share their human resources and historical contexts. Therefore, we introduce a new analysis using the neighboring nations. We consider two distinct definitions of neighboring countries to gain a comprehensive understanding of regionality: i) geographical neighbors and ii) academic neighbors (for details, see Section~\ref{Methods}). Here, the five closest countries are selected as the neighboring countries for a given country.

We first compare the share of the primary country and two types of neighboring groups (Figure~\ref{fig:fig3}). Comparing them reveals that the total proportion of neighboring countries (including the publishing country itself) increases when the primary country is included in the neighboring countries, regardless of whether it is the publishing country (Figure~\ref{fig:fig3} A--B and D--E). 
For journals with identical primary and publishing countries, 66.7\% (76\%) of all journals are dominated by the publishing country and its geographical (academic) neighbors (\textit{i.e.}, have a publication share greater than 60\%).
When the primary country is not the publishing country, only 2.7\% of journals have a primary country among their geographical neighbors, while 70.8\% of journals have a primary country among their academic neighbors. According to these statistics, geographical distance plays no significant role in publishing on a regional scale. However, the relationship between the primary country and neighboring countries changed the publication shares. For journals that neighboring countries include the primary country, a publication share increases by 15.8\% (19.3\%) on average. In contrast, the publication share of the neighboring countries is lower by 24.3\% (14.2\%) than the primary country's publication rate when the primary country is not included in the neighboring countries (Figures~\ref{fig:fig3} C and F).
In short, the profile of neighboring countries may explain the extended regionality of journals, thereby can support the validity of the analysis.

We then apply the concept of neighbors to illustrate the regional characteristics of both questionable and unquestionable journals. Comparing the publication share of geographical neighbors (Figure~\ref{fig:fig2}A and B), unquestioned journals and other journals show similar distributions, which displays a two-peaked distribution with the most frequent or rare publishing; thus, in part, there are regional journals. In contrast, questionable journals lack geographical characteristics. The majority of geographical counterparts of questionable journals rarely publish in the journal. In other words, various countries, not just the geographical neighbors, publish their papers in questionable journals showing their international characteristics. This trend also holds for the academic neighbors (Figure~\ref{fig:fig2}C) with a two-peaked distribution at the maximum and moderate publication share for the unquestionable journals, whereas the distribution for the questionable journals is only single-peaked at the intermediate publication share. Our observation of the primary country's publication share suggests that there are approximately two types of journals: regional journals with a large contribution from the publishing and neighboring countries, and international journals with a negligible contribution from these countries.  

\begin{figure}
    \centering
    \includegraphics[width=\textwidth]{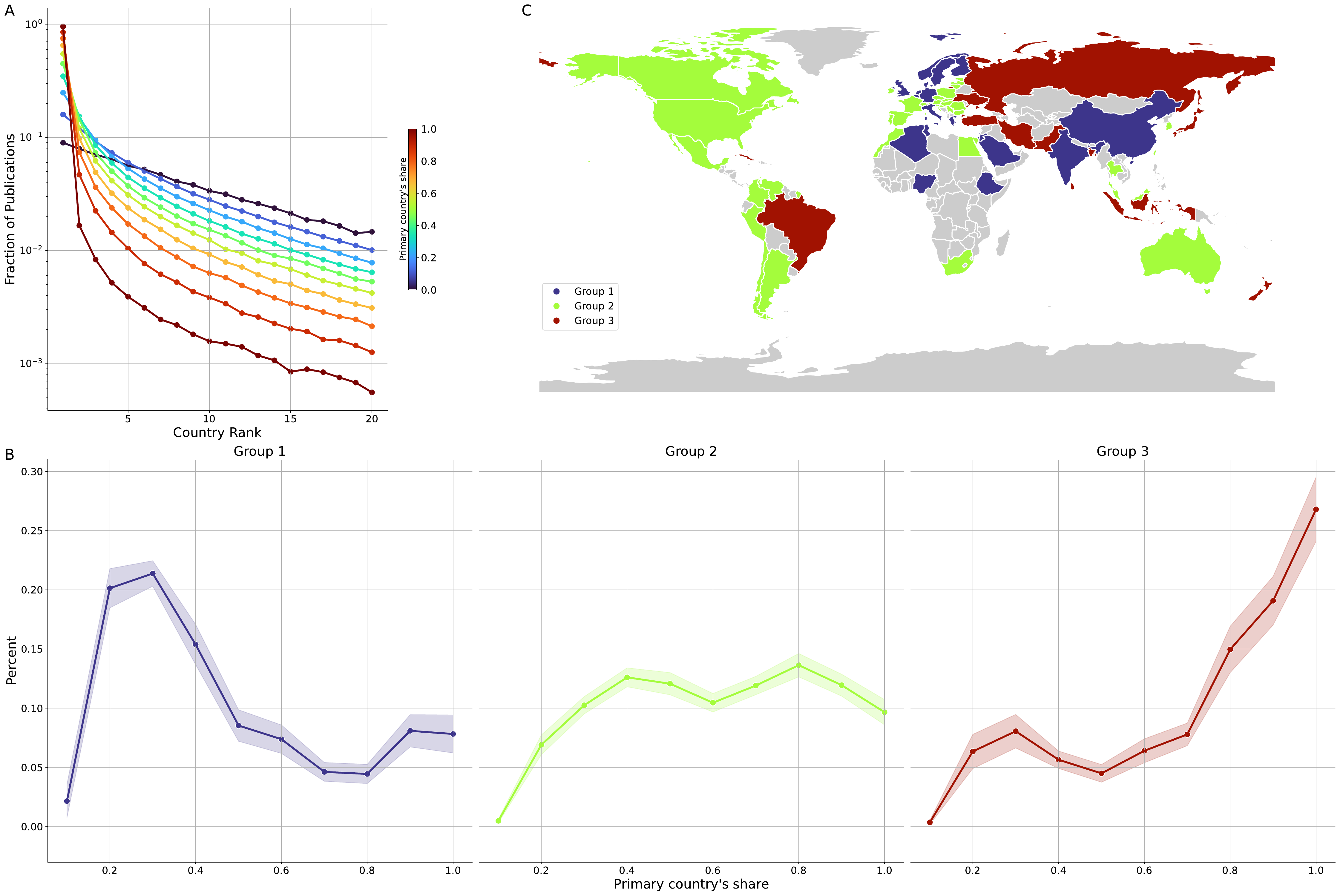}
    \caption{\textbf{The rank-ordered average share of a country's journal publications based on the primary country's publication rate}. 
    \textbf{(A)} The colorbar represents the publication rate of the primary country in the journal. For journals that the primary country's share is high (red), the publication share of other countries is low. For journals in which the primary country's share is low (blue), various countries with lower ranks publish more papers. 
    \textbf{(B)} Publishing preference distribution for each of the three groups. The first group published more in journals with low shares of primary countries (left), while the countries in the second group published comparable percentages in journals with both high and low shares of main countries (center). The third group publishes more in journals that have a large proportion of primary countries (right). The error bars represent the standard error.
    \textbf{(C)} Countries clustered by the distribution of journals' primary countries share profiles. The journals are divided into ten classes based on their share of the primary country in (A) and estimate the portion of each class inside the country (Figure~\ref{fig:sm_country_primary_dist}). We then cluster the countries into three groups based on the estimated portions by Ward's method~\citep{ward1963hierarchical}. Note that we excluded countries with fewer than 10 journals for the analysis.
    }
    \label{fig:fig5}
\end{figure}

A natural step forward is in-depth exploring the share of countries in a given journal. We examined the journal's regional dominance by examining the share of publications by country's rank in a given journal (Figure~\ref{fig:fig5}A). 
The decline of share by rank is slight for journals in that the primary country's publication rate is low. For journals that have the primary country's publication rate of $20-30\%$, the top three countries publish more than $10\%$, and even the twentieth-largest country can produce more than $1\%$ of the journal's total publications. When the primary country's publication rate increases, the publication share is concentrated on the highly ranked countries. For journals that the primary country's publication rate is $80-90\%$, only the primary country produces more than $10\%$ and only the five countries can produce more than $1\%$ of the journal's total publications. Therefore, the publication rate of the primary country can be employed as a proxy for regionality. 

To confirm if the proxy for regionality supports the hypothesis of regional academia, we examine regional dominance, which is a comparative analysis of the selected portion of journals by primary country. Specifically, we calculated the number of journals from each country that is the primary country of the journal (Figure ~\ref{fig:sm_country_primary_dist} for the journal distribution). Using Ward's method~\citep{ward1963hierarchical}, we identified three country groups based on their selection of journals, where 21, 31, and 12 countries consist of each group, respectively. Group 1, comprising countries that publish predominantly in international journals with a low proportion of primary country, was followed by Group 2, comprising balanced publishing countries that publish in journals with a low to high proportion of primary country, and Group 3, comprising countries that publish predominantly in regional journals with a high proportion of primary country (Figure~\ref{fig:fig5}B). In contrast to the classification of many Asian countries as Group 3 (Figure~\ref{fig:fig5}C), European countries are predominantly categorized into Group 1 and Group 2. Group 3 comprises countries with the lower GDP, whereas Group 1 or Group 2 comprises countries with the higher GDP. Group 1 and Group 2 countries have respective mean GDPs $\simeq \$1.6 \times 10^{12}$ and $\simeq \$1.1 \times 10^{12}$, whereas Group 3 countries have an average GDP $\simeq \$5 \times 10^{11}$. The fact that many high-GDP nations are classified as Group 1 or Group 2 suggests that these countries have a greater propensity to publish in international journals. In the academia of these countries, regional journals might not hold significant prominence for authors. In contrast, low-GDP nations depend on regional journals, which suggests the endorsement, support, and even indispensability of regional journals from the regional academia of such countries.

\begin{figure}
    \centering
    \includegraphics[width=\textwidth]{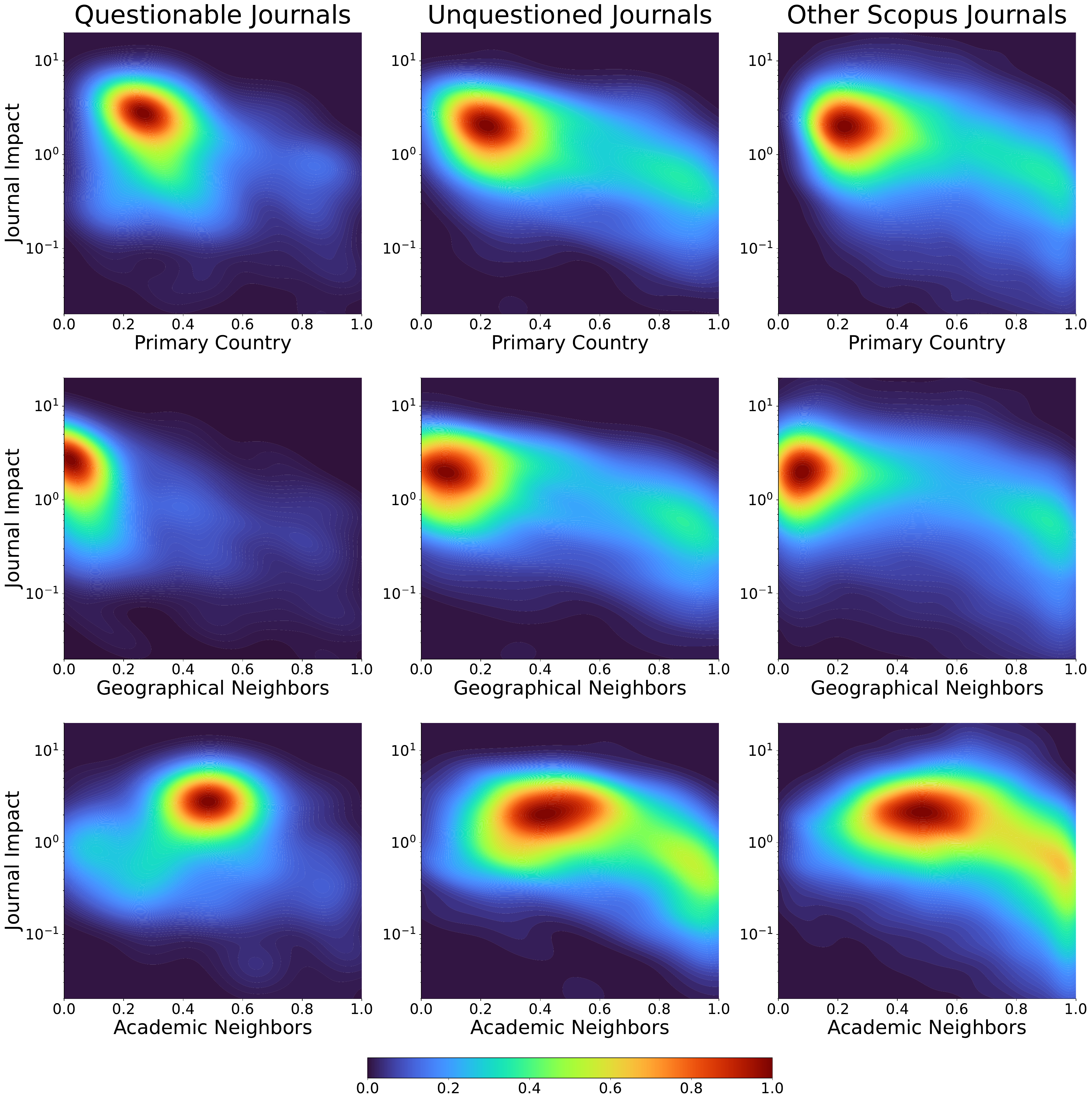}
    \caption{\textbf{Heatmap comparing the primary country's publication share to the journal's impact.} The heatmaps depict the normalized frequency of questionable, unquestionable, and other journals, along with the proportion of the primary country, geographical neighbor, and academic neighbors, as well as the impact of the journal. The red area denotes a dense concentration of journals. The color bar represents the normalized density, which is normalized by maximal density of each plot.}
    \label{fig:fig31}
\end{figure}

A common notion of international journals is their higher impact than regional journals. As a logical step, we measured the share of the primary and neighboring countries to its impact factor by the types of journals (Figure~\ref{fig:fig31}). We observed clear differences between questionable journals and other types of journals. For the unquestioned and other journals, those with a low share of primary and neighboring countries have a higher impact factor than those with a high share, aligning with the common notion. We observe a tail with a gradual decrease in journal impact by the share of the primary and neighboring countries for the unquestioned journals and other Scopus journals, however, questionable journals have a tiny number of journals in those tails. The questionable journals exhibit a low share of the primary and neighboring countries regardless of their journal impact. This observation is valid regardless of the definition of neighbors, which differs questionable from other types of journals. In closing, it is difficult to concur with the hypothesis of questionable journals contributing to regional academia, given the rarity of their function as regional journals.

\section{Discussion}

In this study, we have compared questionable publishing at the journal-country level. The number of questionable publications is proportional to a country's GDP, but the total number of publications in countries with a high GDP conceals its effect on the rise of questionable publications. Compensating the size effect reveals the growing preference for questionable publishing in countries with a high GDP. In contrast to the growing number of dubious publications, they may represent a minority region of academia. However, our analysis reveals that their behavioral patterns differ from the other journals: authors from a greater variety of countries publish their papers in questionable journals, and this behavior is unrelated to geographical and academic collaboration proximity. The location of a questionable journal indicates which country profits from global research funding, while the number of questionable publications in a country indicates who paid for~\citep{eykens2019identifying}. Our analysis of neighboring countries connects the two main agents of publishing. 

We have shown an overestimation of less developed countries to contribute to the increase of questionable journals and publishers. In terms of publication rate, developing countries have a greater volume of questionable publications; yet, because their absolute number of publications is low, their impact on the rise of questionable journals and publishers is limited. This also implies that less developed countries may have been unaware of publishing in questionable journals. Meantime, the country-by-country trend reveals positive correlations with GDP. The comparison demonstrates that questionable publishing is currently gradually permeating developed countries with a high GDP.

Our analysis demonstrates clearly that the questionable journals publish a small proportion of articles from the primary country, as well as from their geographical and academic neighbors. In contrast, the main contributors of unquestioned journals are from the primary and neighboring countries. This trend is also valid for all Scopus journals other than questionable journals. In other words, questionable journals do not have any regional contributions, implying that questionable journals should not be understood as an alternative to resolving the academic dynamics of social stratification.

\subsection{Limitations of the Study}

The list of questionable journals, \textit{i.e.}, \textit{Beall's list}, have biased criteria for selecting questionable publishing; thus, it has inherent limitations for defining criteria. For example, Beall's list is criticized for its negative outlook on open access~\citep {krawczyk2021open}. In addition, there are potentially questionable but non-listed questionable journals, yet we neglected them in the analysis. The publishing country of some questionable journals is purposefully misplaced for various reasons~\citep{demir2018predatory}, but the main result, that is the low percentage of the primary country's publishing, is independent of its declared publication location. Since this study focused on the massive behavior of questionable publishing, individual behavior, \textit{e.g.}, government policy for each country or greedy authors which use \textit{salami slicing} or citation cartel~\citep{kojaku2021detecting}, remains as further study.

We utilized a controlled experiment to examine questionable publishing of unquestioned journals at a comparable position in academia. One clear advantage of a controlled experiment is that one can compensate for the bias from the population. However, one should note that the Scopus dataset cannot cover all scientific publishing; therefore, the comparative analysis, with all Scopus journals, even can have potential bias~\citep{mongeon2016journal}. To surmount such a bias, we also presented the result of all other Scopus journals regardless of their academic positions and found that it shows similar collective behaviors with the unquestioned journals. In addition, to reduce selection bias, we restrict the interpretation to collective behavior rather than individual differences between journals. For instance, when comparing the publication rate of a questionable journal to that of an unquestioned journal, the result may vary depending on the degree of overall regionality of such journals. Consequently, the comparison can distinguish between questionable and unquestionable journals, at least in a collective manner.

\section*{Declaration}

\subsection*{Availability of data and materials}
The data of the list of questionable publishers and journals is public from \url{beallslist.net}. The Socpus data can be requested via Elsevier.

\subsection*{Competing interests}
The authors declare no competing interests.

\subsection*{Funding}
This research was supported by the MSIT (Ministry of Science and ICT), Republic of Korea, under the Innovative Human Resource Development for Local Intellectualization support program (IITP-2022-RS-2022-00156360) supervised by the IITP (Institute for Information \& Communications Technology Planning \& Evaluation). This work was also supported by the National Research Foundation of Korea (NRF) funded by the Korean government (grant No. NRF-2022R1C1C2004277 (T.Y.) and 2022R1A2C1091324 (J.Y.)). The Korea Institute of Science and Technology Information (KISTI) also supported this research with grant No. K-23-L03-C01 (J.P.) and by providing KREONET, a high-speed Internet connection.

\subsection*{Authors' contributions}
All authors designed the experiment, collected the data, and wrote the manuscript. T.Y. analyzed the results. All authors reviewed, edited, and approved the manuscript.

\subsection*{Acknowledgements}
A draft version of this paper was posted on arXiv: \url{https://arxiv.org/abs/2312.07844}

\bibliographystyle{elsarticle-harv} 
\bibliography{main}

\clearpage

\begin{center}
    \item {\fontsize{14}{0}\selectfont \textbf{Supplemental Information for}}
    \item{\fontsize{14}{0}\selectfont Regional profile of questionable publishing}
\end{center}

\setcounter{equation}{0}
\setcounter{figure}{0}
\setcounter{table}{0}
\setcounter{page}{1}
\setcounter{section}{0}

\makeatletter
\renewcommand{\thesection}{Section S\arabic{section}}
\renewcommand{\theequation}{S\arabic{equation}}
\renewcommand{\thefigure}{S\arabic{figure}}
\renewcommand{\figurename}{\textbf{Figure}}

\begin{figure}
    \centering
    \includegraphics[width=0.9\textwidth]{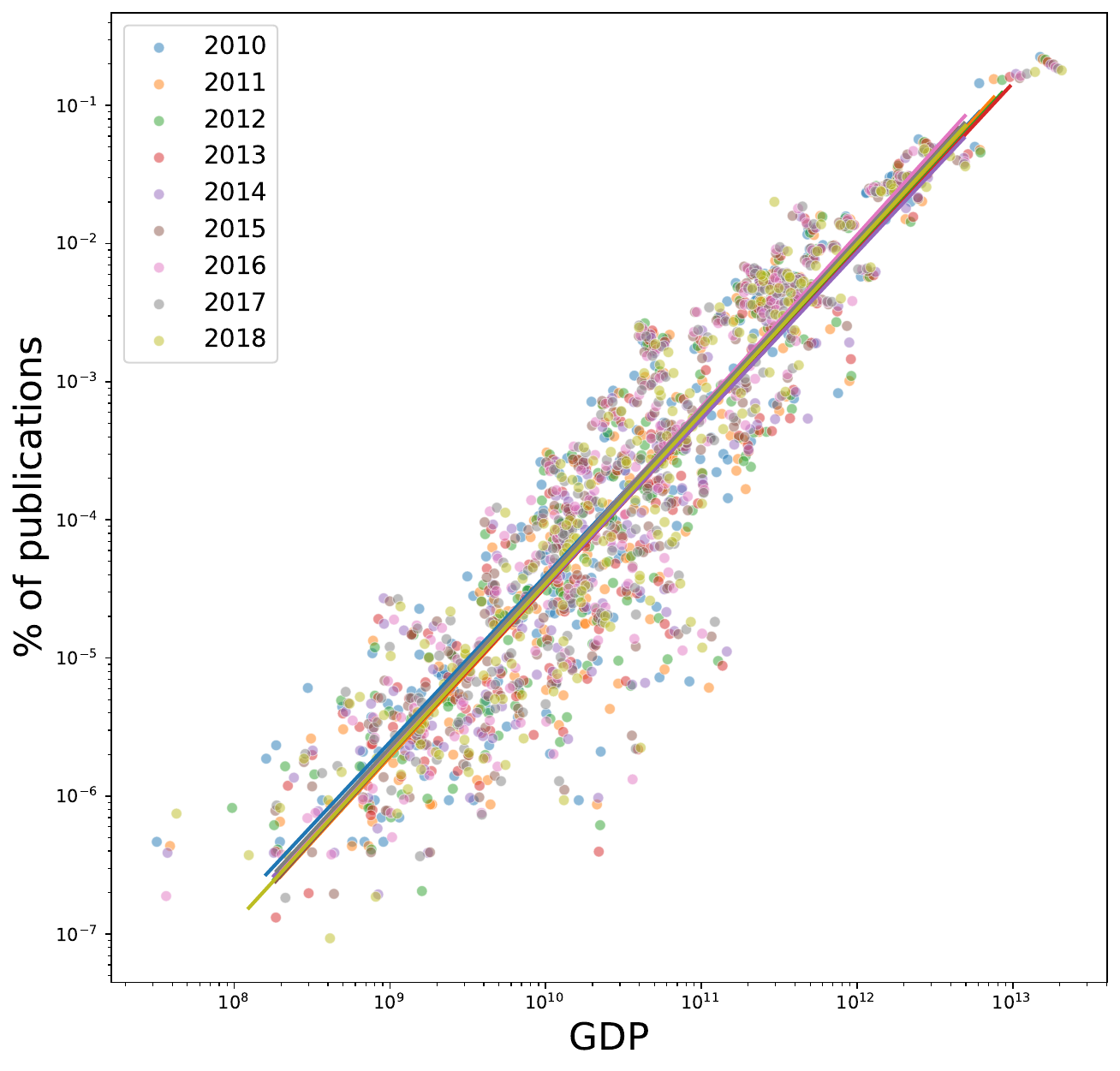}
    \caption{\textbf{The proportion of publications by country compared to GDP}. GDP and proportion of publications are strongly correlated (0.967 in 2010 and 0.972 in 2018). The solid regression line displays the fitted model of $y \sim x^{1.2}$, where $x$ is GDP and $y$ is \% of publications. Top 10 countries produces more than 60\% of publications each year.}
    \label{fig:sm_publication_size}
\end{figure}

\begin{figure}
    \centering
    \includegraphics[width=0.7\textwidth]{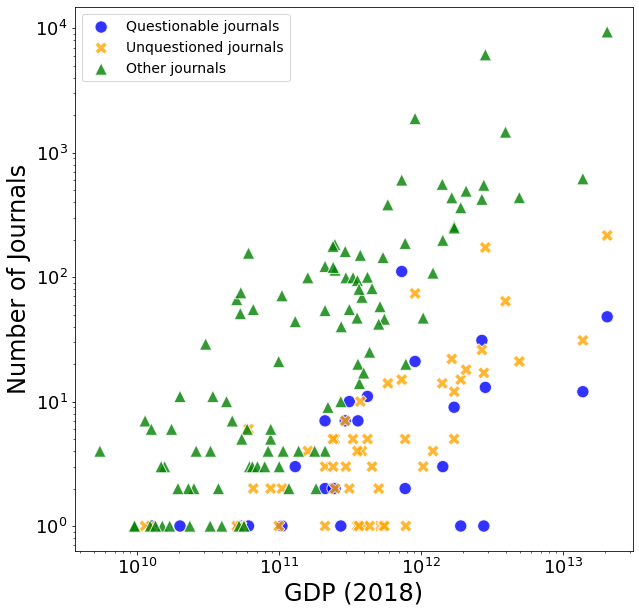}
    \caption{\textbf{The number of journals in the country, plotted by their GDP.} The Pearson correlation between the number of questionable journals and other journals in Scopus is 0.339, while the correlation between unquestioned journals and other journals is 0.987. The Pearson correlation between the number of journals and GDP is 0.30 for questionable journals, 0.72 for unquestioned journals and 0.75 for other journals.}
    \label{fig:sm_journal_size}
\end{figure}

\begin{figure}
    \centering
    \includegraphics[width=0.5\textwidth]{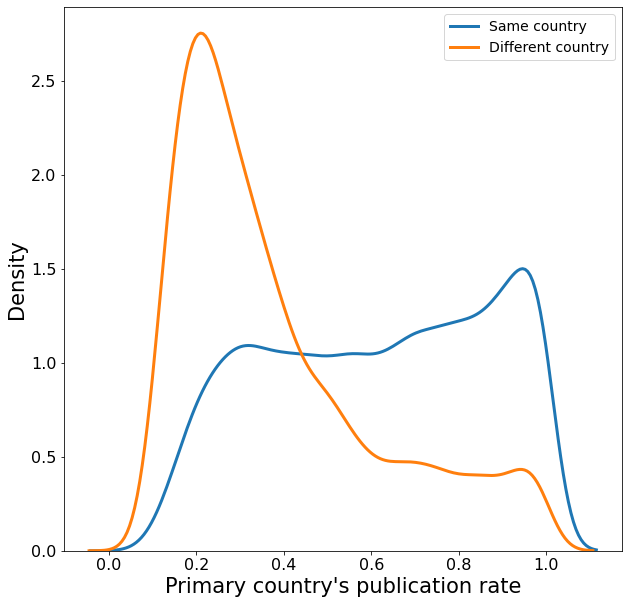}
    \caption{The primary country's publication rate in relation to the publishing country of the journal.}
    \label{fig:sm_dominant_country}
\end{figure}

\begin{figure}
    \centering
    \includegraphics[width=\textwidth]{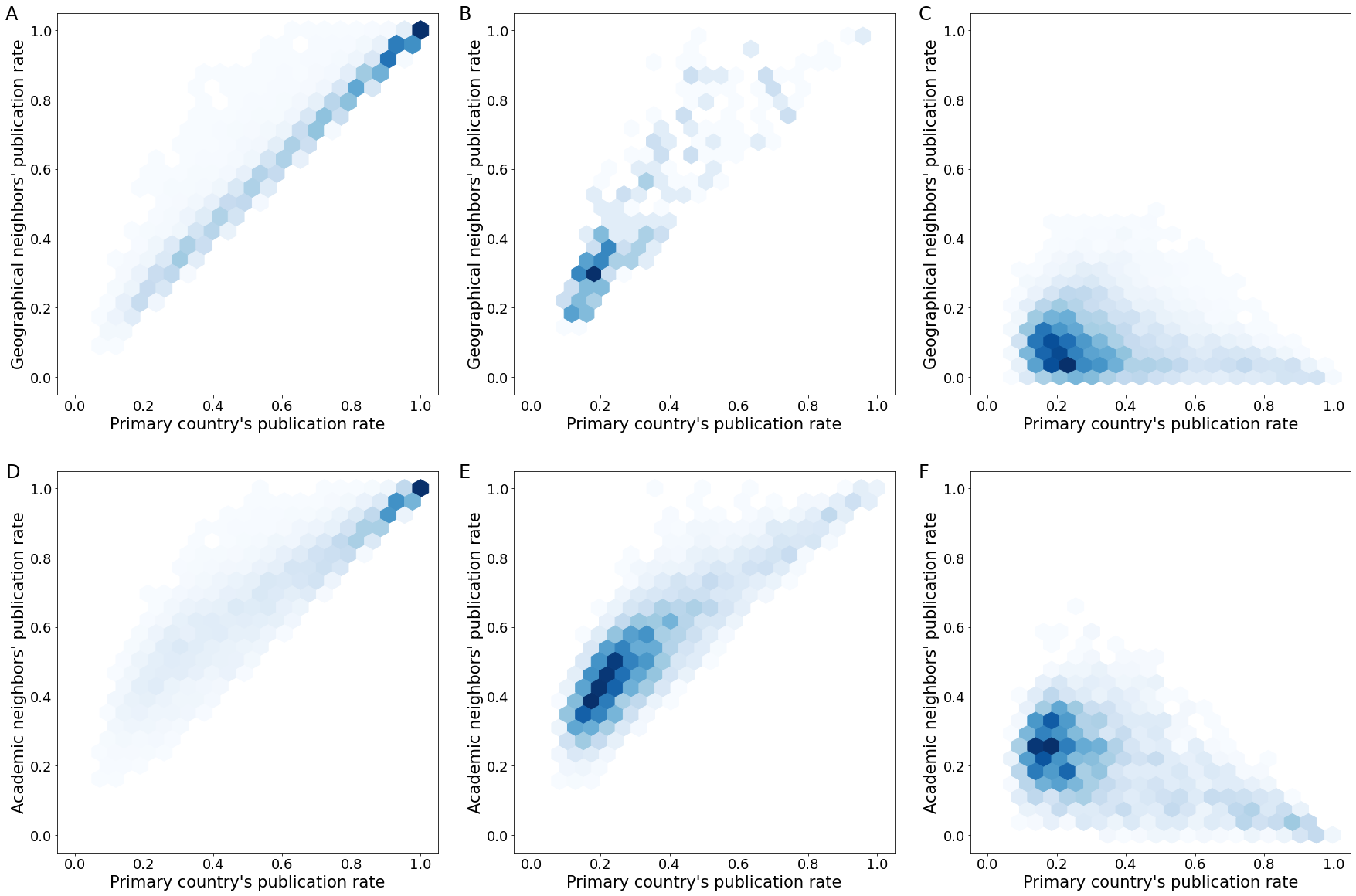}
    \caption{\textbf{Changes of country's proportion with neighbors in the journal.} Two types of neighbors, \textit{i.e.}, (A-C) geographic neighbors and (D-F) co-work neighbors, are used to measure the change of dominant country's proportion of publications, for the case of the two countries are same (A,D) and different (B,C,E,F). When two countries are different, we split whether the dominant country is a neighbor of journal's country (B,E) or not (C,F)
    \textbf{(A)} Geographical neighbors where the dominant country and journal's country are same. 
    \textbf{(B)} Geographical neighbors where the dominant country and journal's country are different, but the dominant country is a neighbor of journal's country.
    \textbf{(C)} Geographical neighbors where the dominant country and journal's country are different, but the dominant country is not a neighbor of journal's country.
    \textbf{(D)} Co-work neighbors where the dominant country and journal's country are same. 
    \textbf{(E)} Co-work neighbors where the dominant country and journal's country are different, but the dominant country is a neighbor of journal's country.
    \textbf{(F)} Co-work neighbors where the dominant country and journal's country are different, but the dominant country is not a neighbor of journal's country.
    }
    \label{fig:fig3}
\end{figure}

\begin{figure}
    \centering
    \includegraphics[width=\textwidth]{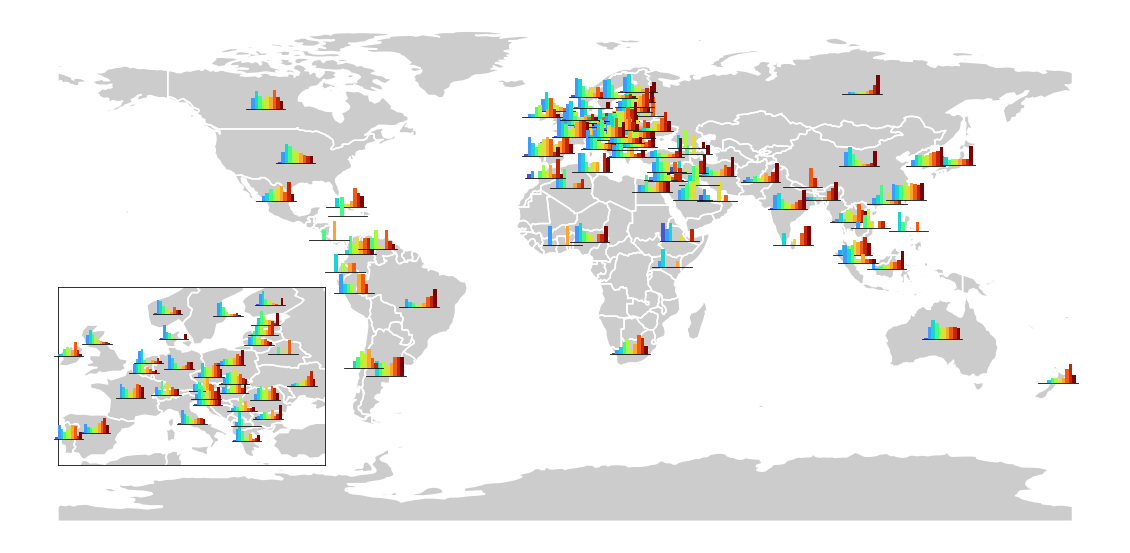}
    \caption{\textbf{Journal distribution of primary country.} The journals are divided into ten groups by the primary country's share of publications. In each country, the proportion of journals is displayed from [0,10]\% (left) to [90,100]\% (right) share of the primary country. The inset shows the journal distribution of European countries.}
    \label{fig:sm_country_primary_dist}
\end{figure}

\noindent\textbf{This PDF file includes:}\\
\\
Supplementary text\\
Figures.~\ref{fig:sm_publication_size} to~\ref{fig:sm_nrcadiff}\\

\newpage

\section{Trend by year}

\begin{figure}
    \centering
    \includegraphics[width=\textwidth]{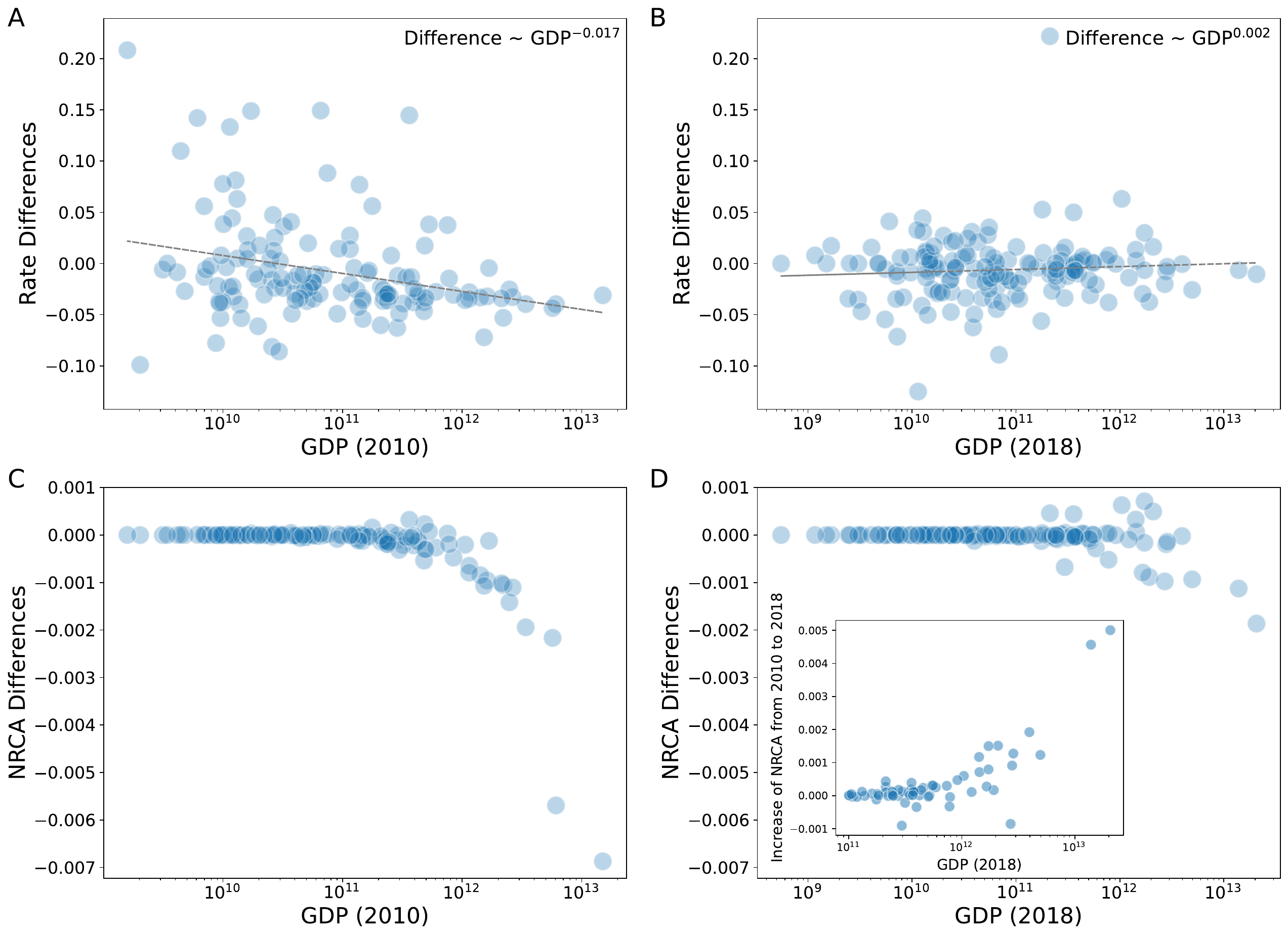}
    \caption{\textbf{The publication trend changes by country.} The difference of publications between questionable journals and unquestioned journals are computed (A,B) by their size and (C,D) by normalized revealed comparative advantage (NRCA). The changes are displayed by year of (A,C) 2010 and (B,D) 2018. In (A,B), fitted line displays the scaling relation between GDP and the differences. See also Figures~\ref{fig:sm_ratediff},~\ref{fig:sm_nrcadiff} for the annual distribution.
    \textbf{(A)} The difference of publication size between questionable journals and unquestioned journals in 2010.
    \textbf{(B)} The difference of publication size between questionable journals and unquestioned journals in 2018.
    \textbf{(C)} The NRCA difference between questionable journals and unquestioned journals in 2010.
    \textbf{(D)} The NRCA difference between questionable journals and unquestioned journals in 2018. Inset displays the increase of NRCA from 2010 to 2018 of the countries that their GDP is higher than $\$10^{11}$}
    \label{fig:fig15}
\end{figure}

\begin{figure}
    \centering
    \includegraphics[width=\textwidth]{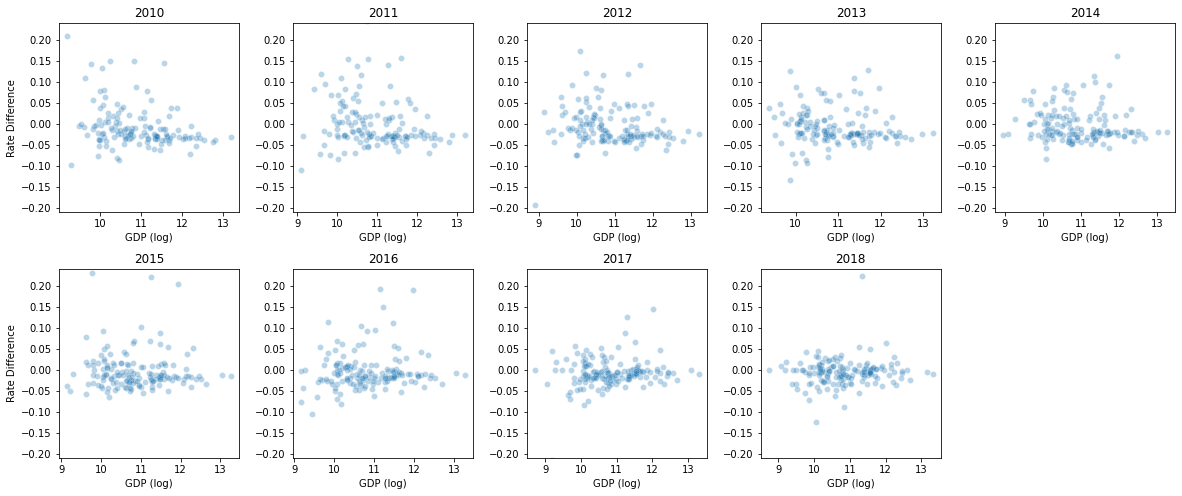}
    \caption{Publication rate difference between questionable publishing and its comparative journals in country.}
    \label{fig:sm_ratediff}
\end{figure}

\begin{figure}
    \centering
    \includegraphics[width=\textwidth]{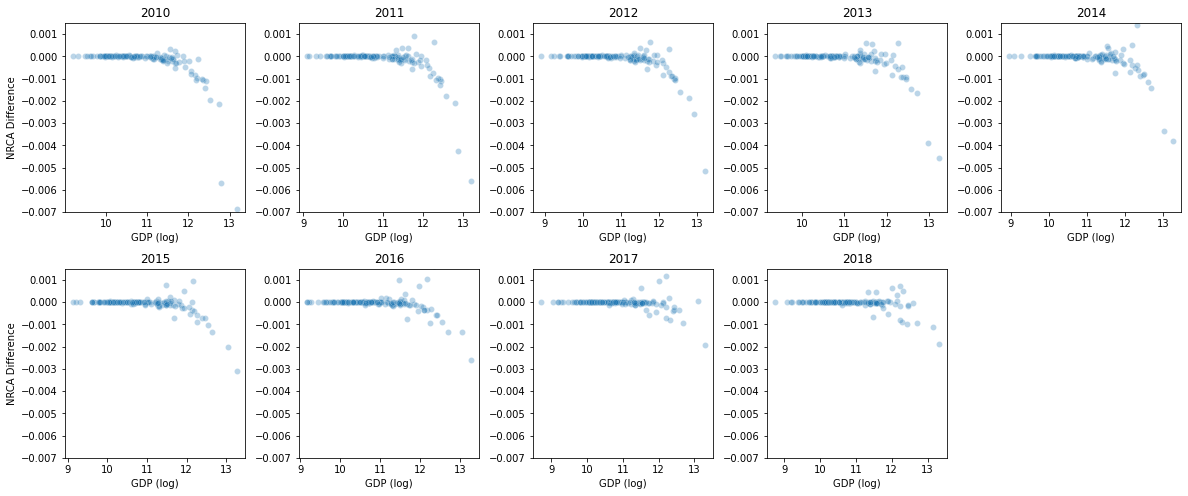}
    \caption{NRCA difference between questionable publishing and its comparative journals in country.}
    \label{fig:sm_nrcadiff}
\end{figure}

The comparison of the publication rate between questionable journals and unquestioned journals exhibits the changes of the publication trend (Figures~\ref{fig:fig15}--\ref{fig:sm_nrcadiff}). In 2010, the difference of publication rate between questionable journals and unquestioned journals shows a negative correlation with GDP (Figure~\ref{fig:fig15}A) with a negative scaling exponent -0.017. Low GDP countries have positive difference, which means they produces more publications in questionable journals rather than other similar journals. High GDP countries shows less difference than low GDP countries. In 2018, the trend changes that a positive correlation with GDP (Figure~\ref{fig:fig15}B) with exponent 0.002. When compared to 2010, low GDP countries produces less questionable publication. Only one country exhibits high number of questionable publications rather than other similar journals, while seven countries show difference more than 0.1.

While the rate does not reflect the number of publications in the country, the Normalized Revealed Comparative Advantage (NRCA) considers both the number of publications in country and in journals~\citep{yu2009normalized}. The NRCA computes the preference of questionable (or unquestioned) publishing of a country compared to the expected number of publications with a randomized assignment. The NRCA is calculated as $\mbox{NRCA}_{j}^i = (R_{j}^i - R_t^i R_{j} / R)/R$, where $R_{j}^i$ denotes the number of publications in group $i$ of country $j$, $R^i$ denotes the sum of publications in group $i$, $R_j$ denotes the sum of publications in country $j$, and $R$ is the number of publications. We assign a group $i$ to questionable, unquestioned, and other journals. We compute the NRCA difference of questionable journals and unquestioned journals by country. When the difference is more (less) than 0, the country prefer to publish in questionable (unquestioned) journals. We found that GDP and the preference are strongly correlated in both 2010 (-0.94) and 2018 (-0.72). The NRCA differences in 2010 show a two types of trends (Figure~\ref{fig:fig15}C). When GDP is lower than $\$10^{11}$, countries shows no preferences. When GDP is higher than $\$10^{11}$, it shows a rapid decrease in NRCA difference. However, NRCA differences increased for high GDP countries in 2018 (Figure~\ref{fig:fig15}D). Inset in Figure~\ref{fig:fig15}D displays the amount of increase in NRCA differences, which clearly shows the preference of questionable publishing grows for high GDP countries. Moreover, we found countries that have a positive value, which means a preference on questionable publishing even though their GDP is high. In conclusion, the change of publication trend support the previous work that publishing in questionable journals are broadening its area from the countries in academic periphery to the academic center~\citep{bagues2019walk,marina2021prevalence}.

\end{document}